\documentclass[conference]{IEEEtran}
\IEEEoverridecommandlockouts

\usepackage{cite}
\usepackage{amsmath,amssymb,amsfonts}
\usepackage{algorithmic}
\usepackage{graphicx}
\usepackage{textcomp}

\usepackage{xcolor}
\usepackage{multirow}
\usepackage{float}
\usepackage{subcaption} 
\def\BibTeX{{\rm B\kern-.05em{\sc i\kern-.025em b}\kern-.08em
    T\kern-.1667em\lower.7ex\hbox{E}\kern-.125emX}}

\usepackage{pifont}

\usepackage{colortbl}
\usepackage{braket}
\usepackage{comment}
\usepackage{balance}
\usepackage{hyperref}

\usepackage{balance}
\usepackage{xcolor}
\usepackage[ruled,vlined,noend]{algorithm2e}

\SetCommentSty{mycommfont}
\SetAlFnt{\small}
\SetAlCapFnt{\small}
\SetAlCapNameFnt{\small}

\usepackage{soul}
\def\BibTeX{{\rm B\kern-.05em{\sc i\kern-.025em b}\kern-.08em
    T\kern-.1667em\lower.7ex\hbox{E}\kern-.125emX}}

\begin{document}

\title{Exploiting Timing Side-Channels in Quantum Circuits Simulation Via ML-Based Methods}
\author{
    \IEEEauthorblockN{ Ben Dong, Hui Feng, Qian Wang}
    \IEEEauthorblockA{Department of Electrical Engineering, University of California, Merced, CA, USA}
    \IEEEauthorblockA{cdong12@ucmerced.edu, hfeng9@ucmerced.edu, qianwang@ucmerced.edu}
}

\maketitle



\begin{abstract}
As quantum computing advances, quantum circuit simulators serve as critical tools to bridge the current gap caused by limited quantum hardware availability. These simulators are typically deployed on cloud platforms, where users submit proprietary circuit designs for simulation. In this work, we demonstrate a novel timing side-channel attack targeting cloud-based quantum simulators. A co-located malicious process can observe fine-grained execution timing patterns to extract sensitive information about concurrently running quantum circuits.
We systematically analyze simulator behavior using the \textbf{QASMBench} benchmark suite, profiling timing and memory characteristics across various circuit executions. Our experimental results show that timing profiles exhibit circuit-dependent patterns that can be effectively classified using pattern recognition techniques, enabling the adversary to infer circuit identities and compromise user confidentiality. We were able to achieve 88\% to 99.9\% identification rate of quantum circuits based on different datasets. This work highlights previously unexplored security risks in quantum simulation environments and calls for stronger isolation mechanisms to protect user workloads.

\end{abstract}

\maketitle

\section{Introduction}

Quantum computing is rapidly advancing, with companies such as IBM and Atom Computing recently announcing processors exceeding one thousand qubits. Despite these developments, access to large-scale fault-tolerant quantum hardware remains prohibitively expensive and limited \cite{shor1996fault, lammers2025quantum}. As a result, quantum circuit simulators have become indispensable for algorithm prototyping, performance benchmarking, and validating quantum applications in the noisy intermediate-scale quantum (NISQ) era. Quantum simulators based on state-vector or tensor-network can scale to dozens of qubits on modern high-performance systems, bridging the gap between theoretical design and hardware deployment \cite{bayraktar2023cuquantum}. Their increasing availability in both open-source packages (e.g., Qiskit Aer, PennyLane Lightning \cite{bergholm2018pennylane}) and commercial cloud services underscores their central role in today’s quantum computing ecosystem. 

However, the growing adoption of simulators on open and cloud platforms raises significant security concerns, as side-channel information can be leaked and exploited to reveal details of quantum circuit designs. Classical cryptographic research has long shown that timing \cite{kocher1996timing, schwarzl2023practical}, memory \cite{wang2023nvleak}, and cache footprints \cite{yarom2014flush+} of the computing systems can betray information about secret inputs. Similar concerns are now emerging in the quantum domain, where prior studies have examined model theft risks in quantum neural networks hosted on untrusted clouds \cite{kundu2024evaluating}, as well as data poisoning vulnerabilities in quantum machine learning pipelines \cite{fu2024quantumleak}. Meanwhile, hardware-level threats such as power side-channels on quantum controllers have also been demonstrated \cite{xu2023exploration}. Yet, little attention has been given to quantum simulators themselves, which, despite being purely classical software, may leak sensitive details about the quantum circuits they execute. Since simulators are widely used by academia and industry and often accessed through shared clusters or third-party cloud providers, adversaries co-located on the same system may infer proprietary details of quantum circuits without direct access to circuit files.

The implications of this study are significant in examining the security properties of quantum simulators. Quantum simulators serve as a staging ground for proprietary algorithms in domains such as quantum chemistry \cite{kassal2010simulating}, finance \cite{matsakos2024quantummontecarlo}, and cryptography \cite{dou2022qpanda}. Leakage of qubit counts, depth, or gate distributions could allow adversaries to reconstruct structural information about circuits, facilitating algorithm fingerprinting or competitive intelligence gathering. Moreover, as simulators are increasingly offered as Simulator-as-a-Service by commercial providers, the attack surface grows beyond local environments into cloud-hosted infrastructure, raising risks analogous to those seen in QNN model theft.
In this work, we present the first systematic study of timing side-channel vulnerabilities in quantum circuit simulators. Using the QASMBench suite \cite{li2023qasmbench} as a representative testbench, we executed both small-scale ($<$10 qubits) and medium-scale (10–27 qubits) circuits on a Linux-based workstation equipped with an NVIDIA RTX 4090 GPU and dual Intel Xeon 6533 processors. During execution, we collected runtime metrics including simulation time, average per-shot latency, and compilation time. 

This paper makes the following key contributions:

\begin{itemize}
   \item We propose the first study of timing side-channels in quantum circuit simulators, demonstrating that execution time can be used to predict qubit counts and gate counts, and further match circuits from the QASMBench dataset.  
   \item Our empirical results show that timing features are strongly correlated with circuit structure, enabling accurate prediction of qubit and gate counts.  
   \item We build a comprehensive dataset of simulation time traces using Qiskit state vector simulator across QASMBench circuits of varying sizes, including small, medium, and combined benchmarks.  

   \item Our findings show that an adversary can predict the number of qubits in a circuit with up to 90\% accuracy and approximate the total gate count with up to 80\% accuracy using timing side-channel observations.  We develop a machine learning–based framework that extracts discriminative features from timing side-channel traces and achieves 88–99\% accuracy in inferring circuit properties.  
\end{itemize}

\section{Background \& Related Work}

\subsection{Quantum Circuit Simulation}
\label{ssec:obfuscation}
\subsubsection{Quantum Gates and Circuits}
Quantum gates and circuits form the foundation of quantum computation, serving as the essential elements that enable the expression and execution of quantum algorithms \cite{divincenzo1998quantum}. 
A qubit differs from a classical bit which can only be 0 or 1. It can exist in a superposition of both 0 and 1 until it is measured, at which point it collapses to either state with probabilities determined by amplitudes $\alpha$ and $\beta$. When multiple qubits interact, they can form entangled states where the qubits are no longer independent and the state of one immediately constrains the state of the other, enabling computational power beyond classical systems.
To manipulate qubits, quantum gates are applied in a manner analogous to classical logic gates \cite{ Barenco1995}. Single-qubit gates such as the Hadamard and Pauli gates operate on individual qubits to create or modify superpositions, while multi-qubit gates such as controlled and phase gates establish entanglement and correlations between qubits. The combination of these gates constructs quantum circuits that provide a structured pathway for implementing quantum algorithms on quantum processors \cite{Nielsen2000}. 



\subsubsection{Quantum Simulator}
On actual hardware, noise and decoherence restrict the number of qubits and circuit depth that can be reliably executed, motivating the heavy use of quantum simulators \cite{Preskill2018}. Quantum simulators provide a classical approximation of quantum state evolution. Among the most widely used are state-vector simulators, which explicitly store the $2^{n}$-dimensional quantum state for $n$ qubits and apply gate transformations through large-scale linear algebra. The complexity of such simulation grows exponentially with qubit count, with each gate operation requiring linear or superlinear time in the state size \cite{Jones2018}. As a result, simulation runtime and memory footprint scale tightly with qubit count, circuit size, depth, and gate composition, making these parameters key determinants of execution behavior.

In practice, simulators such as Qiskit Aer \cite{qiskit-aer}, PennyLane Lightning \cite{bergholm2018pennylane}, and proprietary cloud backends (e.g., AWS Braket SV1) \cite{braket-sv1} serve as critical platforms for algorithm prototyping, benchmarking, and validation. To standardize comparisons, QASMBench has emerged as a representative benchmark suite containing circuits of varying size (from $<10$ to 27 qubits), depth, and gate diversity. Because simulators are implemented in classical software stacks (C++, CUDA, etc.), their runtime is inherently influenced by workload timing characteristics, thus creating an opportunity for side-channel leakage.


\subsection{Timing Side-Channels in Classical and Quantum Domains } 

Side-channel attacks exploit unintentional leakages in physical or software systems. In the classical security domain, timing attacks on cryptographic algorithms such as RSA and AES have shown that input-dependent variations in execution time can reveal secret keys \cite{kocher1996timing, brumley2003remote }. Similar vulnerabilities extend to cache timing and memory access patterns. These attacks succeed not by breaking the underlying algorithm but by exploiting correlations between secret data and the timing signals.

In the quantum domain, side-channel analysis is gaining traction. Researchers have demonstrated power side-channel attacks on superconducting controllers, recovering gate sequences from analog traces \cite{trochatos2023hardware, trochatos2024dynamic, xu2023exploration}. They have investigated and power and fault-based leakages in quantum computing workflows, though primarily at the hardware or algorithm level \cite{Xu2023FaultInjection}. Despite these advances, quantum simulators themselves remain largely ignored in the security literature. Because simulators exhibit deterministic execution dependent on circuit size, depth, and gate structure, they represent a ripe attack surface for adversaries monitoring timing or memory behavior.\\
\noindent\textit{Machine Learning for Side-Channel Inference:}
Recent trends in machine learning have shown that timing and performance traces can be exploited with supervised or unsupervised machine learning to infer sensitive system properties \cite{duddu2018stealing, tizpaz2019efficient, poudel2025machine}. Classical works use decision trees, support vector machines, or deep neural networks to map the side-channel features to key values. In our study, we extend this paradigm to quantum circuit simulators. By collecting multiple timing traces from QASMBench workloads, we demonstrate that qubit count and gate count can be inferred with high accuracy using lightweight ML classifiers, demonstrating the risk of leakage from simulators deployed in untrusted or multi-tenant environments.

\begin{figure*}[t]
    \centering
    \captionsetup{font=normalsize} 
    \includegraphics[width=\textwidth]{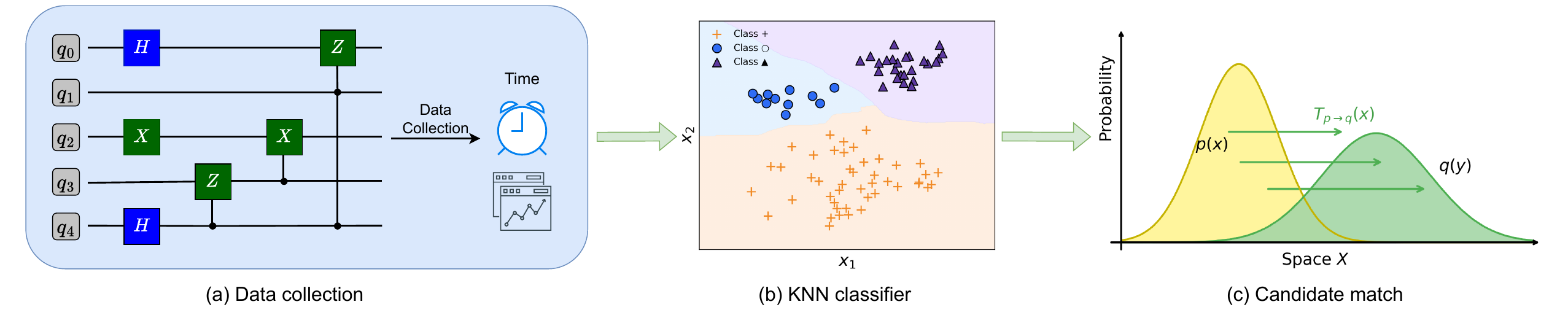}
    \caption{Flowchart of the proposed method: (a) Collect timing data from the quantum circuit. (b) Apply a KNN classifier to predict qubit and gate counts and filter candidates within the feasible range. (c) Compute the Wasserstein distance to match the candidate with a prerecorded circuit.}
    \label{fig:perf_model}
    \vspace{-0.5cm}
\end{figure*}

\subsection{Related Work and Limitations}
Recent studies highlight growing concerns over side-channel attacks in both non-quantum and quantum systems. For example, Erata et al. \cite{erata2024quantum} showed that a single power trace can be used with per-channel brute force or total power mixed integer programming to reconstruct gate-level circuits and reveal secret quantum algorithms. Xu et al. \cite{xu2023exploration} further investigated vulnerabilities in cloud-based quantum computer controllers, introducing several new attack types that enable control pulses and circuits to be reverse-engineered. At the pulse level, other work has shown that mismatches between high-level gate definitions and their actual implementations can be exploited for inference or manipulation \cite{xu2025security}. Beyond quantum controllers, related efforts include taxonomy of hardware Trojans in quantum circuits,\cite{das2024trojan} \cite{conditionaltrojan} logic locking of quantum circuit, \cite{wang2025tetrislockquantumcircuitsplit} \cite{eloq} and methods like SCAUL \cite{ramezanpour2020scaul} and static power analysis that explore unsupervised leakage detection and measurement factors in classical hardware \cite{moos2019static}.

Although these methods show serious risks of side-channel attacks, they rely on access to real quantum hardware and high-quality power measurements that are expensive and often noisy \cite{erata2024quantum}. Collecting data at scale is therefore difficult and error-prone. Our work is the first to explore attacks on a quantum simulator. This makes experiments easier to set up, removes hardware noise, and reduces the cost of evaluation. The simulator also lets us explore many scenarios in a controlled way, which makes our method more practical and scalable while still providing useful insights for real quantum systems.

\section{Threat Models}

In this paper, we treat quantum circuits as unique intellectual property (IP) that faces potential exposure when executed on cloud-based quantum simulators. We focus on adversaries who exploit timing side-channel information to infer details about the quantum circuits. Specifically, we assume that quantum circuits are uploaded to commercial or academic cloud systems for simulation or emulation of quantum phenomena. While the memory containing the circuit description is protected from direct access, similar to traditional assumptions in fault- or power-based side-channel attacks, the execution process may still leak indirect information \cite{xu2023exploration,xu2025security,erata2024quantum}. In particular, the runtime of the simulator becomes a critical leakage channel that adversaries can exploit.

\noindent \textbf{Adversary Capabilities.} We assume that the adversary can monitor the runtime of simulations running on the cloud platform through side channels. By measuring execution times across different circuit inputs, the adversary can build a statistical profile of the computation. Leveraging modern inference techniques, such an attacker can recover approximate circuit characteristics, including the number of qubits involved and the number of gates applied. These properties are sensitive because they reveal the scale, structure, and complexity of proprietary quantum algorithms, which represent valuable intellectual property.

\noindent \textbf{Attack Methodology.} To demonstrate feasibility, we design a machine learning–based inference algorithm that processes timing data and predicts circuit attributes. By correlating runtime behavior with known training examples, the algorithm can accurately estimate qubit counts and gate complexity. This highlights that, even when direct access to circuit descriptions is restricted, adversaries can exploit timing side channels to extract critical IP features from quantum emulation environments.

\section{Methodology}

Our proposed timing side-channel analysis and identification framework consists of three phases as illustrated in Fig.~\ref{fig:perf_model}: 
(i) Data collection, both a summary of data and a per-shot record for each dataset
(ii) Machine Learning model training and analysis using Linear Regression for range prediction and KNN for confidence intervals, and 
(iii) Wasserstein Distance matching to identify the candidate circuit. 
More specifically, given only the timing and memory-related data collected from the \textbf{QASMBench} testing suite, we developed a machine learning based architecture that predicts the underlying circuit's qubit count range, gate count range, and thus conducts the circuit identification. 

\subsection{Stage I: Data Collection}

\subsubsection{Timing Measurement and Memory Profiling}
\begin{figure*}[t]
    \centering
    \captionsetup{font=small} 
    \includegraphics[width=0.95\linewidth]{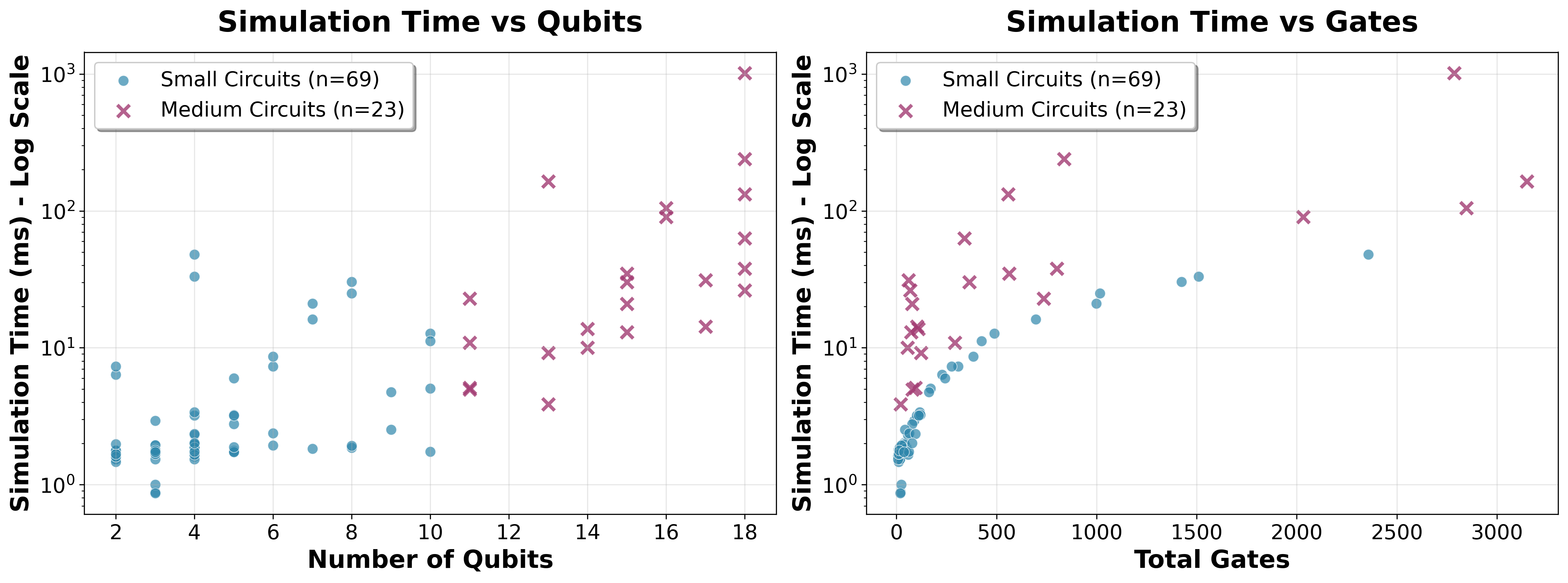}
    \caption{Scatterplot Representation of how each quantum circuit performs in terms of compilation time vs qubit and simulation time vs qubit}
    \label{fig:perf_model2}
    \vspace{-0.4cm}
\end{figure*}

Execution traces are obtained by running QASMBench circuits on a state-vector simulator under controlled conditions. Both small-scale  ($2$--$10$ qubits) and medium-scale ($11$--$27$ qubits) benchmarks are included. Each circuit is executed repeatedly for 100 runs, and warm-up shots are excluded.  As shown in Fig.~\ref{fig:perf_model2}, we can already see a pattern in timing-derived data, which scales with qubit gate count.

For each run, the framework records average, median, minimum, and maximum shot times, as well as standard deviation and variance of timing. Memory usage is monitored in parallel, capturing both average and maximum memory deltas. Outlier removal ensures that collected profiles are statistically consistent.  

\subsubsection{Feature Vector Construction}

From the collected data, a multi-dimensional feature vector is constructed per circuit. The selected features include:
\begin{itemize}
    \item \texttt{avg\_shot\_time\_ns}: mean latency per shot,
    \item \texttt{median\_shot\_time\_ns}: median shot latency,
    \item \texttt{min\_shot\_time\_ns} and \texttt{max\_shot\_time\_ns},
    \item \texttt{std\_shot\_time\_ns} and \texttt{timing\_variance},
    \item \texttt{avg\_memory\_delta\_bytes}
    \item \texttt{max\_memory\_delta\_bytes}.
\end{itemize}
These features are standardized via $z$-score normalization across 
the combined dataset:
\begin{equation}
z = \frac{f - \mu_f}{\sigma_f},
\end{equation}
where $\mu_f$ and $\sigma_f$ are the column mean and standard deviation. Columns with $\sigma_f = 0$ are scaled with $\sigma_f = 1$ to preserve stability.

\subsection{Stage II: Candidate Range Prediction and Filtering}

\subsubsection{KNN-Based Range Estimation}
The first stage of candidate filtering relies on $k$-nearest neighbor (KNN) search applied in the standardized feature space. Each circuit is represented by a feature vector $\mathbf{x} \in \mathbb{R}^d$, where the dimensions correspond to timing and memory statistics. After $z$-score normalization, all features are unitless and comparable in magnitude, preventing bias from scale differences between timing (nanoseconds) and memory (bytes).

Given a target vector $\mathbf{x}_t$, we compute the Euclidean distance to all other circuits:
\begin{equation}
d_i = \|\mathbf{x}_t - \mathbf{x}_i\|_2, \quad i=1,\dots,N,
\end{equation}
and identify the $k=5$ closest neighbors with the smallest $d_i$. Each neighbor has an associated ground-truth label $y_i$, representing either 
the number of qubits ($n_{qubits}$) or the total number of gates 
($total\_gates$). Rather than predicting a single point estimate, the 
framework computes an interval spanning the neighbor labels:
\begin{equation}
[lo, hi] = [\min_{j \in N_k} y_j,\; \max_{j \in N_k} y_j],
\end{equation}
where $N_k$ is the index set of the $k$ nearest neighbors. This produces 
a conservative prediction range that captures the plausible values for 
the target circuit. Intuitively, this step exploits the correlation between 
timing/memory patterns and circuit size: circuits with similar side-channel profiles tend to have comparable structural parameters.  

Two separate applications of this procedure are performed: one in 
qubit-labeled space to yield $[lo_{qubits}, hi_{qubits}]$, and one in 
gate-labeled space to yield $[lo_{gates}, hi_{gates}]$. These ranges are 
then used to constrain the set of candidate circuits considered in later 
stages.

\subsubsection{Intersection Filtering}
The second stage applies the predicted ranges as hard filters. First, all circuits whose qubit count lies outside the predicted interval $[lo_{qubits}, hi_{qubits}]$ are eliminated. This already removes a large fraction of the benchmark, while retaining very high coverage of the true circuit. Next, the remaining candidates are filtered again using the gate-count range $[lo_{gates}, hi_{gates}]$. Only circuits that satisfy 
\begin{align}
lo_{qubits} &\leq n_{qubits} \leq hi_{qubits}, \label{eq:qubit_range} \\
lo_{gates}  &\leq total\_gates \leq hi_{gates}. \label{eq:gate_range}
\end{align}

are carried forward.  

The final candidate set is defined as the intersection of both constraints. In practice, this reduces the search space from the full QASMBench dataset to approximately $10$–$14$ circuits on average. While the ranges may be wide—particularly for medium-scale circuits where gate counts vary by hundreds—the intersection strategy ensures that the true circuit is retained with near-perfect probability (97--100\% coverage in experiments). 
By narrowing the space of plausible circuits, this stage makes subsequent distribution-level matching computationally tractable and adversarially feasible.

\subsection{Phase III: Statistical Analysis and Final Candidate Selection}

\subsubsection{Wasserstein Distance Matching}
\begin{figure}[H]
\vspace{-0.3cm}
    \centering
    \captionsetup{font=small} 
    \includegraphics[width=1.0\linewidth]{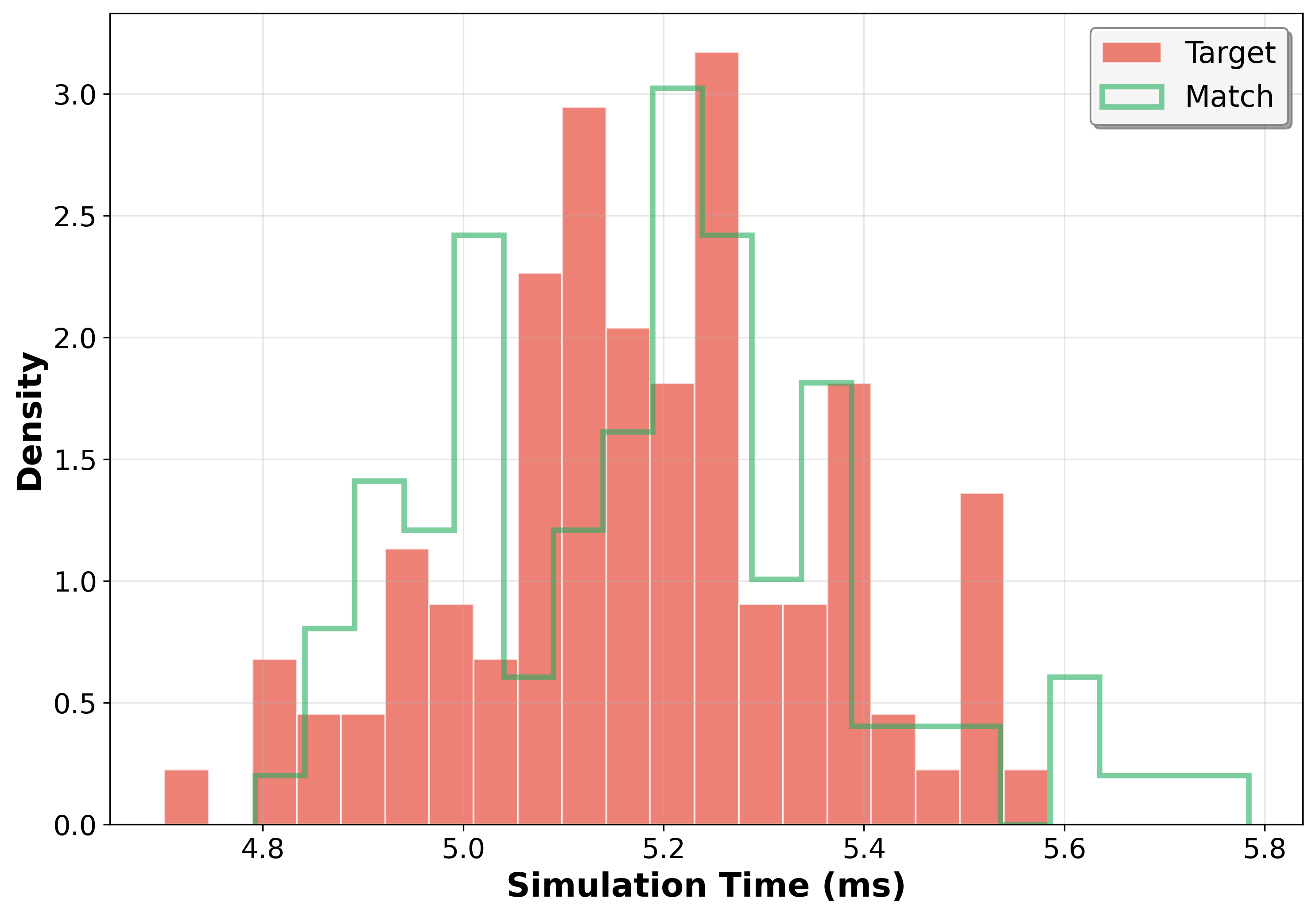}
    \caption{Illustration of Wasserstein Distance.}
    \label{fig:wasserstein_illustration}
\end{figure}
The reduced candidate set is further distinguished using per-shot timing distributions. For each candidate, the one-dimensional Wasserstein distance is computed against the target distribution:
\begin{equation}
W(P,Q) = \inf_{\gamma \in \Gamma(P,Q)} \int |x-y| d\gamma(x,y),
\end{equation}
where $P$ and $Q$ are empirical timing distributions.  
Candidates are ranked by ascending $W(P,Q)$, and Top-1 accuracy is 
reported for evaluation.  \\
As illustrated in Fig.~\ref{fig:wasserstein_illustration}, the method effectively matches the distribution of target circuit and candidate circuit. Even small Wasserstein distances ( are sufficient to provide discriminative power, highlighting why distribution-level matching is essential after KNN range filtering.

\subsubsection{Side-Channel Security Implications}
By combining range filtering with distribution-level matching, the framework demonstrates that timing and memory features alone enable accurate inference of structural properties. The narrowing effect of the KNN ranges ensures that the true circuit remains within the candidate set in almost all cases, while Wasserstein matching provides fine-grained discrimination to identify the correct circuit. This pipeline models a realistic adversary operating under restricted observation channels, highlighting the practical risk of simulator-based side-channel leakage.  

\medskip
In summary, our methodology integrates systematic circuit parsing, 
high-resolution timing collection, machine learning inference, and 
security assessment to expose the vulnerability of quantum simulators 
to timing side-channel attacks.

\section{Experiments and Evaluation}

\subsection{Experimental Setup}

All experiments were conducted on benchmark circuits from QASMBench, 
covering both small-scale ($2$--$10$ qubits) and medium-scale 
($11$--$27$ qubits) circuits. Circuit-level summary data were 
drawn from enhanced timing CSVs, and per-shot execution traces 
were obtained from reference datasets. A total of $92$ circuits 
($69$ small and $23$ medium) were successfully evaluated.  

Each circuit is represented by timing and memory feature vectors 
(average, median, minimum, and maximum shot time; standard deviation 
and variance of timing; average and maximum memory deltas). All 
features are $z$-score normalized across the combined dataset.  
Candidate filtering is performed using $k=5$ nearest neighbor search 
followed by range intersection, and the final identification is conducted 
via one-dimensional Wasserstein distance matching on per-shot timing 
distributions.
\vspace{-0.23cm}
\subsection{Metrics}
We evaluate the framework using the following metrics:
\begin{itemize}
    \item \textbf{Range coverage:} fraction of circuits whose true 
    qubit or gate label lies within the predicted KNN range.
    \item \textbf{Range width:} average size of the predicted qubit 
    and gate intervals.
    \item \textbf{Final candidate size:} number of circuits remaining 
    after intersection filtering.
    \item \textbf{Top-1 accuracy:} probability that the true circuit 
    is ranked first by Wasserstein distance among the final candidates.
\end{itemize}

\begin{table}[h]
\centering
\caption{Classification performance for predicting Qubit and Gate Count.}
\begin{tabular}{l|c|c}
\hline
\textbf{Algorithm} & \textbf{Accuracy} & \textbf{CV Accuracy ($\pm$ std)} \\
\hline
\multicolumn{3}{c}{\textbf{Qubit Classification}} \\
\hline
MLP Neural Net & 77.27\% & 51.24\% $\pm$ 10.41\% \\
Gradient Boosting & 72.73\% & 66.27\% $\pm$ 4.37\% \\
Random Forest & 68.18\% & 59.28\% $\pm$ 15.37\% \\
\hline
\multicolumn{3}{c}{\textbf{Gate Count Classification}} \\
\hline
MLP Neural Net & 72.73\% & 71.90\% $\pm$ 9.64\% \\
Gradient Boosting & 63.64\% & 71.90\% $\pm$ 12.73\% \\
Random Forest & 72.73\% & 70.85\% $\pm$ 8.52\% \\
\hline
\end{tabular}
\label{tab:ml_results_split}
\vspace{-0.3cm}
\end{table}

\subsection{Overall Results}
\begin{figure}[htb!]
    \centering
    \captionsetup{font=small} 
    \includegraphics[width=1.0\linewidth]{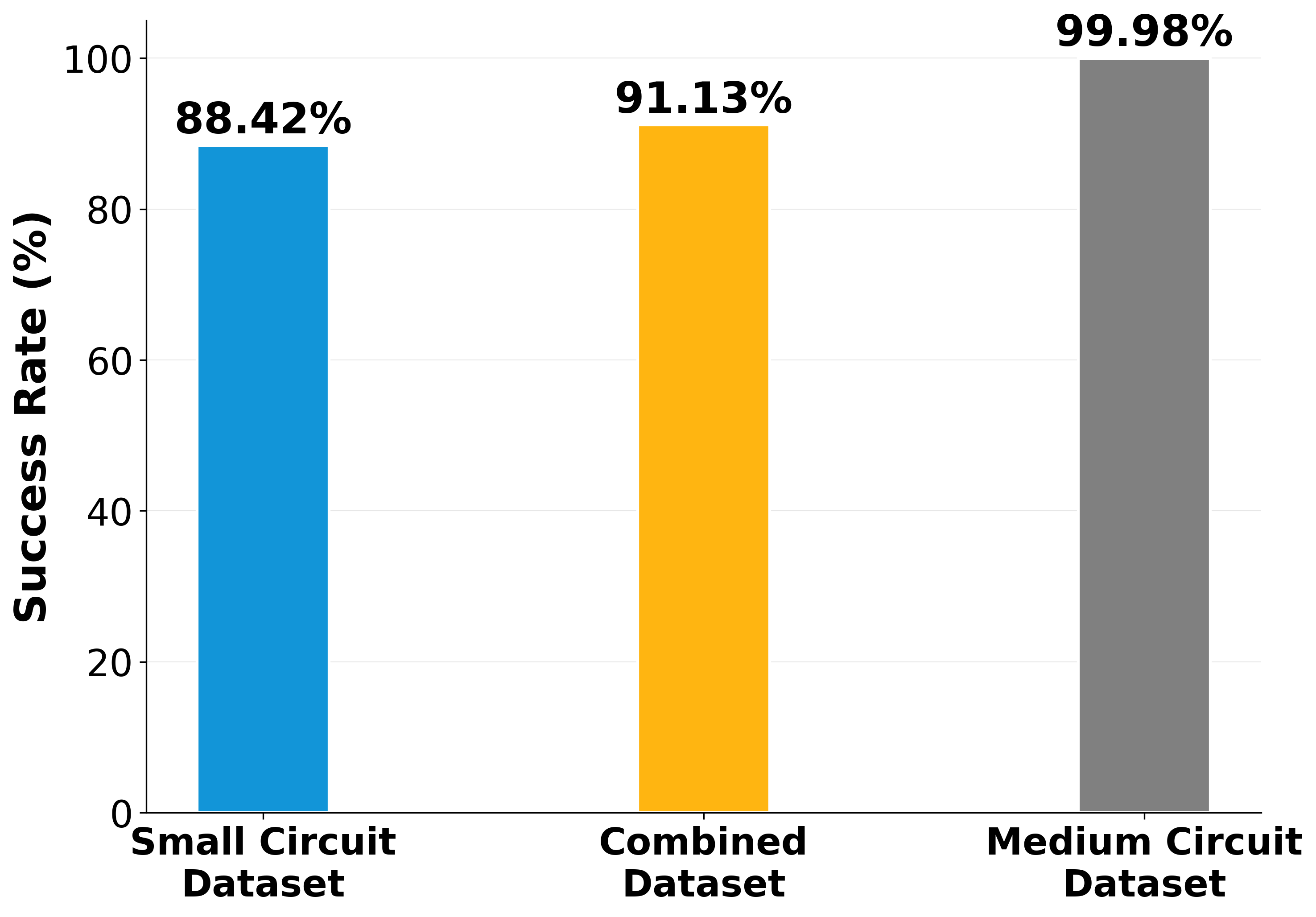}
    \caption{Identification Success Rate Based on Datasets Tested.}
    \label{fig:perf_model1}
    \vspace{-0.5cm}
\end{figure}
Table~\ref{tab:ml_results_split} presents the performance of baseline machine learning classifiers for predicting qubit and gate counts. While MLP neural networks achieve the highest raw accuracy for qubits (77.27\%) and gates (72.73\%), their cross-validation performance drops significantly, with standard deviations exceeding 10\%. Gradient Boosting and Random Forest show slightly more stable results but remain below 72\% average accuracy. These findings indicate that direct classification on timing and memory features is unreliable for precise prediction.

In contrast, our two-step framework achieves far higher success rates by combining KNN-based range estimation with distribution-level Wasserstein matching. As shown in Fig.~\ref{fig:perf_model1}, the overall Top-1 identification accuracy reaches 91.13\% across the combined dataset. When stratified by circuit scale, success varies: small circuits achieve 88.42\% accuracy, while medium circuits reach nearly perfect identification at 99.98\%. This demonstrates the effectiveness of narrowing the candidate set before final matching, and highlights the scale-dependent nature of timing leakage.


\subsection{Small vs. Medium Circuits}
For \textbf{small circuits} ($2$–$10$ qubits), as shown in the Fig.~\ref{fig:perf_model1} the final Top-1
identification rate reached $88.42\%$. All observed failures occurred in this
small-circuit regime. The primary failure mode was inaccurate gate-range
prediction, where the true gate count fell just
outside the estimated interval. A secondary failure mode was qubit-range
mis-classification. These errors stem from the limited dynamic range
of small-circuit timing data, high relative sensitivity to measurement noise,
and dense packing of many circuits in a narrow feature space, which reduces the
discriminative power of both KNN ranges and Wasserstein distance.  

For \textbf{medium circuits} ($11$–$18$ qubits), per the Fig.~\ref{fig:perf_model1} our method achieved a near perfect
Top-1 identification rate of $99.98\%$. This improvement arises from the stronger
signal-to-noise ratio in medium circuits: longer execution times and larger
memory deltas yield distinctive timing fingerprints that overshadow measurement
noise. The sparser distribution of medium-scale circuits in feature space also
reduces ambiguity, making range estimation and distribution matching more
effective.  

Taken together, these results highlight a scale-dependent vulnerability: small
circuits are harder to classify due to overlapping and noisy timing profiles,
while medium circuits are uniquely identifiable with near certainty. As circuit
sizes grow, timing signatures become increasingly distinct, suggesting that the
security risk posed by timing side-channels escalates rather than diminishes at
larger scales.

\subsection{Per-Circuit Observations}
Analysis of individual circuits reveals that circuits with simple 
structures (e.g., \texttt{adder\_n4}, \texttt{bb84\_n8}) are 
consistently identified with zero Wasserstein distance to their 
reference distributions, yielding near $100\%$ Top-1 accuracy.  
Transpiled variants with higher gate counts (e.g., 
\texttt{adder\_n10\_transpiled}, \texttt{basis\_trotter\_n4\_transpiled}) 
are also reliably identified despite broader gate ranges, because their timing distributions remain distinct.  

Misidentifications occur in cases where timing profiles overlap 
significantly across different circuit families. For example, 
\texttt{basis\_change\_n3\_transpiled} was occasionally confused with 
\texttt{pea\_n5}, and \texttt{variational\_n4\_transpiled} was mapped 
to its untranspiled counterpart. These errors correspond to very small 
Wasserstein distances between structurally different but 
timing-similar circuits, illustrating the limits of distribution-only 
matching.

\subsection{Contextual Comparison}

For context, KNN-only prediction without Wasserstein matching yields 
much weaker discrimination: small-circuit qubit-only KNN has point 
accuracy of $26.1\%$ and MAE of $1.8$ qubits, while medium-circuit 
qubit-only KNN has point accuracy of $34.8\%$ and MAE of $1.9$ qubits.  
This highlights the importance of the two-step intersection filtering 
and distribution-level matching: together they raise identification 
accuracy to nearly $90\%$ while keeping candidate sets tractable 
($\approx 10$–$14$ circuits).

\subsection{Discussion}

These results confirm that timing and memory side-channels can be 
exploited to extract structural information from quantum simulators.  
Even without gate-derived features, the adversary can reduce 
the candidate set from nearly one hundred circuits to fewer than fifteen, 
and correctly identify the true circuit in nearly nine out of ten cases.  
The narrowing effect of the KNN ranges ensures near-perfect coverage, 
while the Wasserstein matching provides the fine discrimination needed 
to resolve candidates.  

This demonstrates both the feasibility and the severity of timing-based 
side-channel attacks in quantum computing environments. As circuit sizes 
increase, timing signals become even more distinct, suggesting that 
leakage will persist unless mitigated by countermeasures such as 
timing randomization, constant-time simulation kernels, or memory padding.

Future work will extend this framework in three directions: (i) scaling experiments to larger circuits and tensor-network simulators, (ii) integrating adversarial machine learning models for stronger inference attacks, and (iii) evaluating practical defenses and quantifying their cost-performance tradeoffs. Ultimately, safeguarding quantum simulators against timing side-channels is a necessary step toward ensuring the confidentiality and trustworthiness of quantum computing in both academic and commercial cloud environments.

\vspace{-0.1cm}
\section{Conclusion}

This work presented the first systematic study of timing side-channel leakage in quantum circuit simulators. Using the QASMBench benchmark suite on a state-vector simulator, we showed that execution-time and memory-only features provide enough signal to infer structural properties of quantum circuits. Our framework integrates $k$-nearest neighbor (KNN) range estimation, intersection filtering, and Wasserstein distance matching, enabling accurate recovery of qubit and gate counts as well as circuit identification. Across 92 circuits, qubit ranges achieved 97–100\% coverage and gate ranges exceeded 91\%, with final candidate sets reduced to 10–14 circuits. Within this narrowed space, distribution matching achieved perfect Top-1 accuracy on medium-scale circuits and over 88\% on small circuits, confirming that simulator timing can leak sensitive information.

Our analysis further revealed scale-dependent risks: small circuits are harder to distinguish due to noisy and overlapping traces, while medium circuits yield uniquely identifiable signatures. This counterintuitive finding indicates that larger quantum workloads are even more vulnerable to timing attacks, raising significant security concerns for cloud-hosted simulators. Unless mitigated, adversaries could exploit these leaks for algorithm fingerprinting, competitive intelligence, or reconstruction of proprietary designs. To address this, we recommend countermeasures such as randomized timing, constant-time kernels, workload batching, and memory padding to reduce leakage.

\bibliographystyle{IEEEtran.bst}
\bibliography{qref}

\end{document}